\documentstyle[11pt,paspconf,epsf]{article}

\begin{document}

\title{The view of galaxy formation from Hawaii: Seeing the dark side of
    the universe.}

\author{L.L. Cowie and A.J. Barger}

\affil{Institute for Astronomy, University of Hawaii,
    2680 Woodlawn Drive, Honolulu, HI 96822}



\begin{abstract}
The strength of the submillimeter background light shows directly that much of
the energy radiated by star formation and AGN is moved to far infrared
wavelengths. However, it is only as this background at 850 $\mu$m has been
resolved with direct submillimeter imaging that we have seen that it is
created by a population of ultraluminous (or near ultraluminous) infrared
galaxies (ULIGs) which appear to lie at relatively high redshifts ($z>1$).
Mapping the redshift evolution of this major portion of the universal star
formation has been difficult because of the poor submillimeter spatial
resolution, but this difficulty can be overcome by using extremely deep cm
continuum radio observations to obtain precise astrometric information since
the bulk of the brighter submillimeter sources have detectable radio
counterparts. However, with this precise position information available, we
find that most of the submillimeter sources are extremely faint in the optical
and near infrared ($I>>24$ and $K=21-22$) and inaccessible to optical
spectroscopy. Rough photometric redshift estimates can be made from combined
radio and submillimeter energy distributions. We shall refer to this procedure
as millimetric redshift estimation to distinguish it from photometric
estimators in the optical and near IR. These estimators place the bulk of the
submillimeter population at $z=1-3$, where it corresponds to the high redshift
tail of the faint cm radio population. While still preliminary, the results
suggest that the submillimeter population appears to dominate the star
formation in this redshift range by almost an order of magnitude over the
mostly distinct populations selected in the optical-ultraviolet.
\end{abstract}



\section{Introduction}

The cosmic far infrared (FIR) and submillimeter (SMM) background, which is the
cumulative rest frame FIR emission from all objects lying beyond our Galaxy,
has recently been detected by the {\it FIRAS} and {\it DIRBE} experiments on
the {\it COBE} satellite (Puget et al.\ 1996; Guiderdoni et al.\ 1997;
Schlegel et al.\ 1998; Fixsen et al.\ 1998; Hauser et al.\ 1998) and found to
be comparable to the total unobscured emission at optical/UV wavelengths.
This result shows directly that much of the energy released by the totality of
star formation and AGN radiation through the lifetime of the universe has been
dust absorbed and reradiated into the rest-frame FIR. This in turn implies
that, to obtain a full accounting of the history of the universal star
formation, we must turn our attention to this dark side of the universe.

\section{Resolving the submillimeter background}

 The first stage in this process is to locate the individual
objects giving rise to the background.
Resolution of the extragalactic submm background at 850 $\mu$m
became possible almost simultaneously with the measurement
of the background when 
the Submillimeter Common User Bolometer Array (SCUBA;
Holland et al.\ 1999) was installed on the
15-m James Clerk Maxwell Telescope
(JCMT) on Mauna Kea. SCUBA's
sensitivity and area coverage enabled the
sources producing the submm background to be directly imaged
for the first time. The current count determinations
determined from blank field surveys and from cluster lensed
fields 
(Smail, Ivison \& Blain 1997;
Barger et al.\ 1998, 1999b;
Hughes et al.\ 1998;
Blain et al.\ 1999;
Eales et al.\ 1999) are shown in Figure 1. Barger, Cowie and
\begin{figure}
\plotone{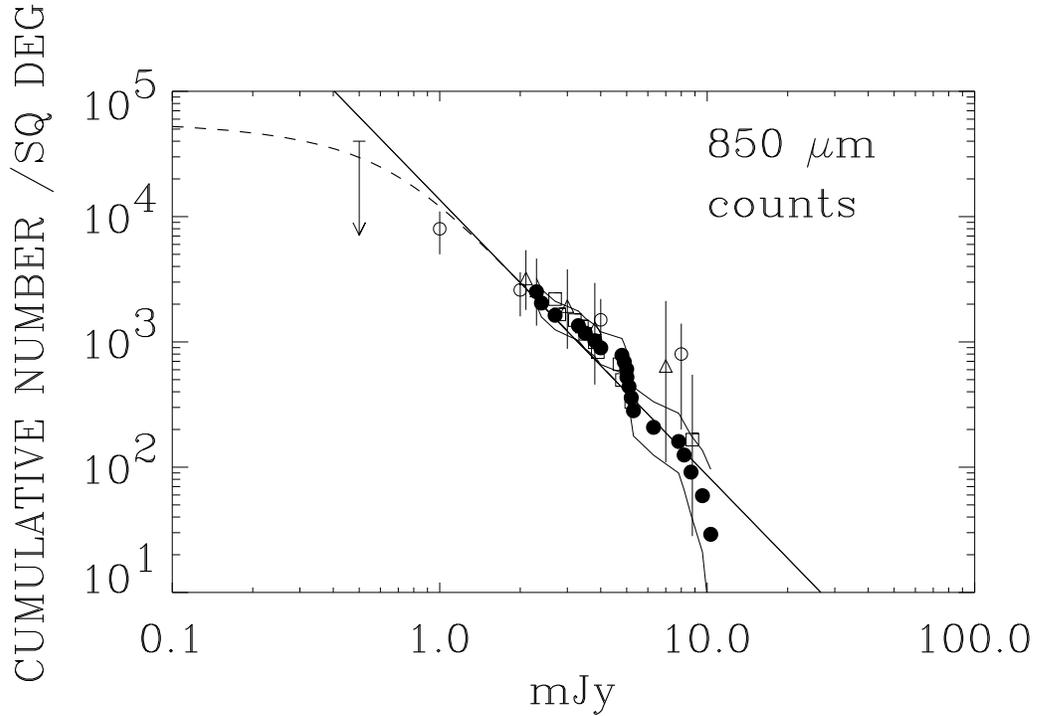}
\caption{The 850 $\mu$m source counts from Barger, Cowie, and Sanders 1999
(solid circles) with $1\sigma$ error limits
(jagged solid lines) are well described by the power-law
parameterization in Eq.~1 with $a=0.4-1.0$, $\alpha=3.2$, and
$N_0=3.0\times 10^4$\ deg$^{-2}$\ mJy$^{-1}$ (solid line).
The dashed curve shows a smooth extrapolation of this fit to match the
EBL measurements
using the value $a=0.5$. Counts from Blain et al.\ (1999)
(open circles), Hughes et al.\ (1998) (open triangles), and
Eales et al.\ (1999) (open squares) are in good agreement with our data
and the empirical fit. 
} \label{fig-1}
\end{figure}
Sanders have shown, using optimal fitting techniques combined with Monte Carlo
simulations of the completeness of the count determinations, that the
cumulative counts are well fit by a power law above 2 mJy.
In addition, they showed that, in
order to match the background and fit to the very limited (one source!)
information at fainter magnitudes from the lensed sample (Blain et al.\ 1999),
a differential source count law
\begin{equation}
n(S)=N_0/(a+S^{3.2})
\end{equation}
was reasonable. Here $S$ is the flux in mJy,
$N_0=3.0\times 10^4$ per square degree per mJy, and $a=0.4-1.0$ is chosen
to match the 850 $\mu$m extragalactic background light. The 95 percent
confidence range for the power law index is from 2.6 to 3.9.
The extrapolation  suggests that the typical SMM source producing the
bulk of the background lies
at around 1 mJy, and the direct counts show that roughly 30 percent of the 
850 $\mu$m background comes from
sources above 2 mJy.  

Provided only that the redshifts lie near or above $z=1$ (see below), the
far-infrared (FIR) luminosity is approximately independent of the redshift.
Thus, if we assume an Arp~220-like spectrum with $T=47$\ K (e.g., Barger et al.\
1998), the FIR luminosity of a characteristic $\sim 1$\ mJy source is in the
range $4-5\times 10^{11}\ {\rm h_{65}^{-2}\ L_\odot}$ for a $q_0=0.5$
cosmology ($7-15\times 10^{11}$ for $q_0=0.02$).  The FIR luminosity provides
a measure of the current star formation rate (SFR) of massive stars (Scoville
\& Young 1983; Thronson \& Telesco 1986), ${\rm SFR}\sim 1.5\ \times 10^{-10}\
(L_{FIR}/{\rm L_\odot})\ {\rm M_\odot\ yr^{-1}}$; a 1\ mJy source would
therefore have a star formation rate of $\sim 70\ {\rm h_{65}^{-2}\ M_\odot\
yr^{-1}}$\ for $q_0=0.5$, placing the `typical' submm source at or above the
high end of extinction-corrected SFRs in optically-selected galaxies (Pettini
et al.\ 1998). If we were to allow the dust temperature to go as low as 30\ K,
$L_{FIR}$ and the corresponding SFR would be $\sim 4$ times smaller.

\section{Direct attempts at a redshift distribution}

The identification of the optical/near-infrared counterparts to
the SCUBA sources is made difficult by the uncertainty in
the 850 $\mu$m SCUBA positions and by the intrinsic faintness of
the counterparts. Barger et al.\ (1999c)
presented a spectroscopic survey of possible optical counterparts
to a flux-limited sample of galaxies selected from the 850 $\mu$m
survey of massive lensing clusters by
Smail et al.\ (1997, 1998).
The advantage of a lensed
survey is that the clusters magnify any background sources,
thereby providing otherwise unachievable sensitivity in the submm,
and easing spectroscopic follow-up in the optical.
In the Barger et al.\ survey, identifications were attempted for
all objects in the SCUBA error-boxes that were bright enough for
reliable spectroscopy; redshifts or limits were obtained for
24 possible counterparts to a complete sample of 16 SCUBA sources.
The redshift survey produced reliable identifications for
six of the submm sources:
two high redshift galaxy pairs (a $z=2.8$ AGN/starburst pair
(Ivison et al.\ 1998) and a $z=2.6$ Lyman-break--like pair (Ivison
et al.\ (1999)),
two galaxies showing AGN signatures ($z=1.16$ and $z=1.06$),
and two cD galaxies (cluster contamination).
The galaxy pairs were later confirmed as the true counterparts
through the detection at their redshifts of CO emission in the millimeter
(Frayer et al.\ 1998, 1999). Because AGN are very uncommon in optically
selected spectroscopic samples, it is also probable that the AGN
identifications are correct, and they place a rough lower limit of
about 20 percent on the fraction of the submillimeter sources which
have AGN characteristics. These results suggest that, excluding
the cluster objects, about a quarter of the submillimeter sources
can be spectroscopically identified.

  However, two of the submillimeter sources in the sample have no counterparts
to $I\/$ around 26 and, while the remaining eight sources have optical galaxies
within the large error circles, these are rather normal objects which may
simply be chance projections.  We shall show in the next section that this is
very probably the case. These missing sources may be at higher redshifts, or
be more dust obscured, than the spectroscopically identifiable sources.
Furthermore Smail et al.\ (1999), using deep near IR and optical imaging of the
fields, have recently detected two extremely red objects which may be the
counterparts of two of these sources, rather than the nearby bright spiral
galaxies which Barger et al.\ observed as the most likely counterparts to the
SMM sources. This result also suggests that many of the optical source
identifications may be suspect.

\section{Positional determination from cm continuum radio observations}

     To proceed further we need accurate astrometric positions,
and these are most easily obtained using cm radio continuum
observations. Because of the well known radio-FIR correlation,
both the cm data and the submillimeter observations are linearly
dependent on the star formation rates in the galaxy (Condon 1992),
though the ratio of the 850 $\mu$m flux to the cm radio flux
rises rapidly as a function of redshift because of the opposite
signs of the K-correction in the two wavelength ranges. (We discuss
this further in the next section.) Because of the redshift dependence,
a cm flux-limited sample will contain a high proportion of
lower redshift objects, while the 850 $\mu$m sample will choose out
primarily the high redshift objects.

\begin{figure}
\plotone{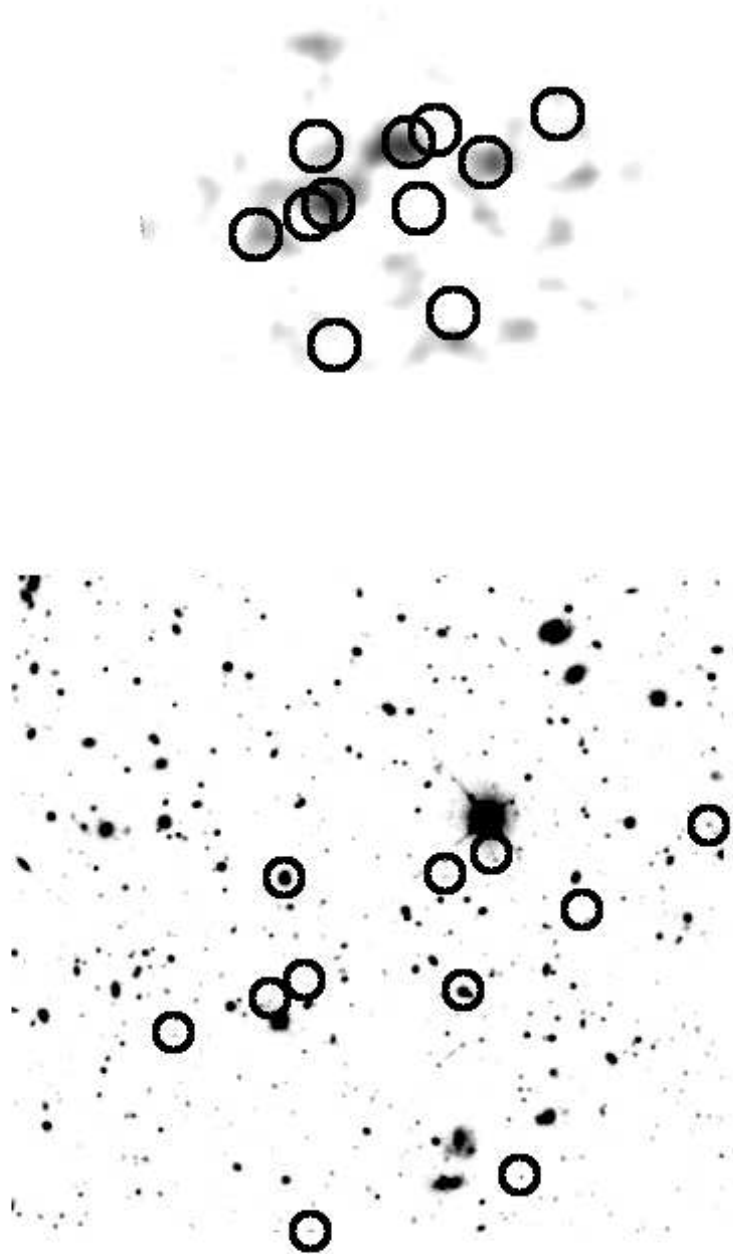}
\vspace{-0.7in}
\caption{Radio sources in the HFF: the figure shows an
overlay of the 20 cm radio sources in a small region
of the HFF on a SCUBA 850 $\mu$m image on the 
top and on a near IR image of the region on the
bottom. (N is to the right and E to the top in these
images.) In general it is the radio sources which are faint
in the optical and near IR that are detected in the submillimeter.} \label{fig-2}
\end{figure}

    The flanking field region of the Hubble deep field (the HFF) is well
suited to looking at the radio versus submillimeter selection.  Eric Richards
(1999) has recently obtained an extremely deep VLA 20 cm image of this region,
with a relatively uniform ($1\sigma = 8 \mu$Jy) sensitivity over the whole
flanking field region, which can be combined with the deep ground based
optical and near IR (NIR) imaging of the HFF (Barger et al.\ 1999a). Richards
et al.\ (1999c) find that roughly two thirds of the $5\sigma$-selected 20 cm
population have relatively bright optical/NIR counterparts while the remaining
third are very faint.  Barger, Cowie, and Richards (1999) have observed a
complete subsample of the radio-selected objects with the LRIS on the Keck II
10m telescope, and find that nearly all the objects with $K < 20$ can be
spectroscopically identified, with a maximum redshift of around 1.2; however,
almost none of the fainter objects were identified.

   From a total sample of 70 radio selected galaxies in the HFF region,
Barger et al.\ (1999c)
chose the 16 with $K > 21$ for followup with SCUBA. However, because
they used the jiggle map mode which provides approximately a 5 square
arcminute field around the target, a large fraction of the remaining
radio sources (35/54) were also serendipitously measured. 
14 of the 16 targeted blank field
sources were observed. Even with relatively shallow SCUBA observations
($3\sigma=6$ mJy at 850 $\mu$m) a very large fraction of the blank
field radio sources were detected in the submillimeter, as is illustrated 
in Figure 2. Of the 14 targeted sources, 5 are detected above 6 mJy
while by contrast none of the 35 optical/NIR bright sources were detected.
In the observed fields, which covered slightly more than half of the
HFF, a further two sources brighter than 6 mJy were discovered which were not in the
radio sample. Even if there are further non-radio--detected submillimeter
sources at the same level in the remaining unobserved portions of the
HFF, it appears that the radio selection is turning up the majority
of the bright submillimeter sources.

\begin{figure}
\plotone{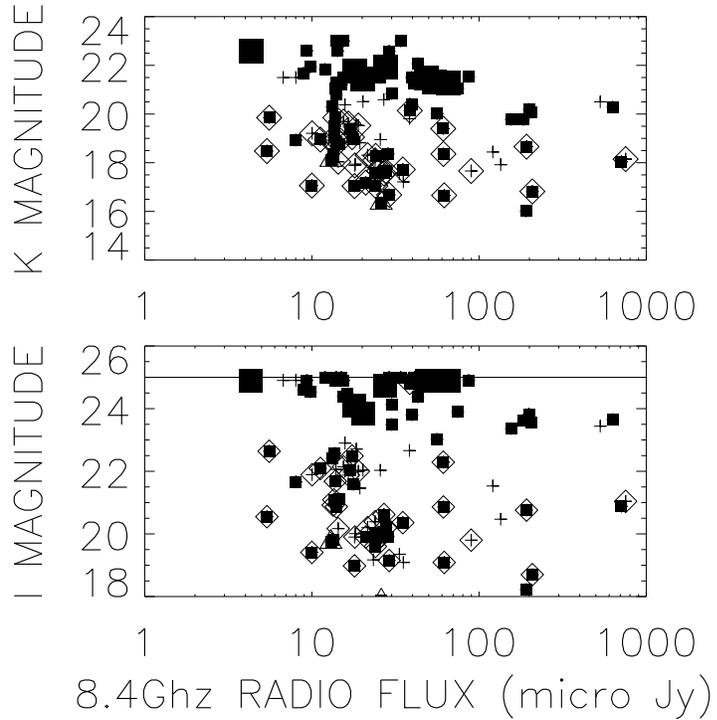}
\caption{The optical and NIR magnitudes of the radio sources 
in the HFF and the SSA13 field versus 8.4GHz flux: the whole
radio population is shown as crosses with sources which only
have 20 cm fluxes extrapolated to 8.4GHz assuming a synchrotron
spectrum. Sources with known redshifts are shown with open
diamonds and all lie at $z<1.2$ except for two quasars in the
SSA13 field (Windhorst et al.\ 1995) which are shown as triangles.
Sources which have been observed in the  850 $\mu$m but not
detected at the typical 6 mJy ($3\sigma$) level are shown as the small
squares while those detected at 850 $\mu$m are shown as the
large squares.
} \label{fig-3}
\end{figure}

   The fact that many of the bright submillimeter sources
can be identified with the optical/NIR faint radio sources
in this way has the extremely important corollary that
many of the 850 $\mu$m-selected sources have extremely
faint optical/NIR counterparts. This is illustrated in
Figure 3, where we show the $K\/$ and $I$ magnitudes of 
radio-selected sources in the HFF, and also in 
the SSA13 field (Richards
et al.\ 1999c and references therein) where a similar 
submillimeter survey has been carried out (Cowie, Barger,
and Richards 1999). Extremely deep $K\/$ observations
with NIRC on the Keck I 10m can yield detections of nearly
all the 850 $\mu$m detected radio sources, and these
are found to 
lie in the $K=21-22$ range.  However, 
many of the sources are not
detected in the $I\/$ band at the $2\sigma$ limit of $I=25$ for
the HFF, and, for the HDF850.2 source in the HDF proper
(Hughes et al.\ 1998), the source is not seen to $I=29$. 
The brightest sources lie in the $I=24-25$ range.

\section{Millimetric redshift estimation}
    While it is clear from the work described in section 3
that a fraction of the submillimeter sources have optical
and NIR counterparts that are bright enough for spectroscopic
identification, the results of section 4 show that a very large
fraction simply cannot be identified in this way. At the current
time the small numbers of objects suggest that perhaps a quarter
of the sources (of which a fairly large fraction have AGN
characteristics) are bright in the optical and spectroscopically
identifiable, while the remainder fall into the optical/near
IR faint category. For this latter category of objects we
will have to rely on photometric estimates using the shape
of the spectral energy distribution in the radio and submillimeter,
and the submillimeter to NIR ratios (Carrilli and Yun 1999, 
Blain et al.\ 1999).

Carrilli and Yun have 
suggested the use of the  850 $\mu$m to 20 cm
flux ratio as such a redshift indicator. Because of the opposing
spectral slopes of the synchrotron spectrum in the radio and
the black body spectrum in the submillimeter, the submillimeter
to radio ratio rises extremely rapidly with redshift, as is shown
in Figure 4,  which is take from Barger, Cowie, and Richards (1999)
where a much more extensive discussion may be found. The primary
uncertainty in this quantity lies in the dust temperature
dependence, which in the local ULIG sample produces a range in 
the ratio of about a multiplicative factor of 2 relative to
Arp220. 

\begin{figure}
\plotone{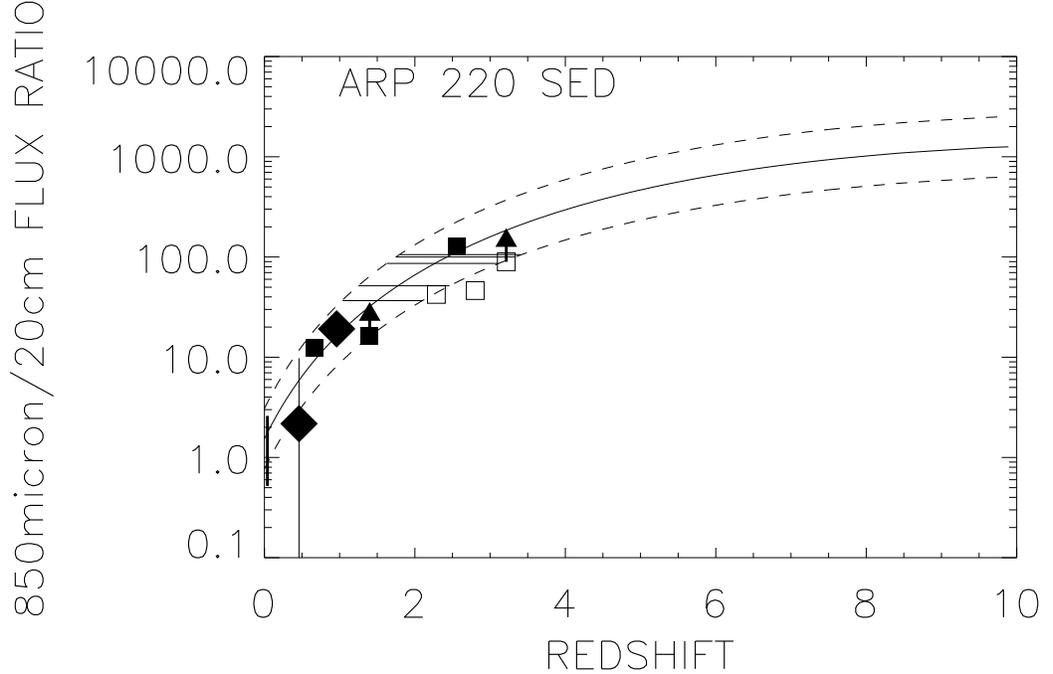}
\caption{Millimetric redshift estimation: The solid curve shows
the ratio of the 850 $\mu$m to 20 cm flux that a non evolving
ARP220 would have as a function of redshift. The solid bar
at low redshift shows the range of the 850 $\mu$m fluxes to
the 8.4GHz flux at low redshift extrapolated to 20 cm assuming
a synchrotron spectrum with the data taken from Rigopoulou et
al.\ (1996). This suggest that the ULIGS have a range of about 
a multiplicative factor of 2 relative to ARP220 which is shown
by the dashed lines. The average submillimeter to 20 cm ratio
of the galaxies which have spectroscopic redshifts and also
have been observed in the submillimeter are shown as the
large diamonds with $1\sigma$ errors. The lowest point is consistent
with a null detection but in the higher redshift bin there is
a strong positive detection consistent with an ARP220 ratio.
Individually detected objects from Lilly et al.\ (1999) and Ivison
et al.\ (1999) are also consistent within the error spread though
those with AGN characteristics (open symbols) appear to fall
low in the figure. The solid line show the best guess for the
redshift range (typically $z=1-3$) for the optically/NIR faint
galaxies which are detected in the submillimeter. Radio objects
which are not detected in the submillimeter are likely to lie
at lower redshifts than this while submillimeter detected objects
which are not seen in the radio are potentially at higher redshift.
} \label{fig-4}
\end{figure}

   We can test the estimator in a variety of ways. In Figure
4 we have shown the average submillimeter to radio ratios for
the objects in the HFF with known spectroscopic redshifts.
While none of these sources is individually detected, the
average values are consistent with a null result at low
redshifts but a strongly significant positive detection for
the sources near $z=1$, which is extremely consistent with
the Arp220 ratio. Individual submillimeter sources with
spectroscopic identifications are also broadly consistent with the expected
ratios, though there is a suggestion that, as might be expected,
those with the AGN characteristics have slightly lower ratios,a
though still within
the broad general range.

   The optically/near IR faint radio sources in the HFF with
submillimeter detections are shown as the horizontal lines in
the figure. The redshift estimator places them in the same broad
general $z=1-3$ range as the typical spectroscopically identified
sources. (For AGN we may be systematically underestimating the
redshifts). Radio sources without  850 $\mu$m detections probably
lie at lower redshifts while the  850 $\mu$m sources without
radio counterparts may represent the high end redshift tail.

\section{Discussion}

We can summarise the results as follows. Roughly 30 percent of the  850 $\mu$m
background is already resolved and the slope of the counts is
sufficiently steep (a power law index of --2.2 for the cumulative
counts) that only a small extrapolation to fainter
fluxes will result in convergence to the background. Thus, the typical
submillimeter source contributing to the background seems to be
in the 1-2 mJy range. Because of the direct correspondence between
flux and luminosity at these wavelengths we may identify the sources
with ULIGs or near ULIGs. About a quarter of the sources have optical
counterparts which are bright enough to be
spectroscopically identified, and a large fraction of these show
AGN characteristics, though at least one is an extremely bright pair
of Lyman break galaxies. However, many of the remaining sources are
extremely faint in the optical and NIR ($K=21-22$). Redshift
estimation for these sources using the submillimeter to radio
ratios places the bulk of them in the
same $z=1-3$ range as the spectroscopically identified sources. 

It is interesting to consider where this population fits in the
overall history of the universal star formation.  One uncertainty
in doing this is the question of what fraction of the SMM light
is powered by AGN rather than star formation.
It has long been debated whether the dust-enshrouded local ULIGs are
powered by massive bursts of star formation induced by violent
galaxy-galaxy collisions or by AGN activity. A recent mid-infrared
spectroscopic survey of 15 ULIGs by
Genzel et al.\ (1998) found that $70-80$\ per cent
of the sample are predominantly powered by star formation and
$20-30$\ per cent by a central AGN.
Thus, while the spectroscopic follow-up studies of the gravitationally
lensed submm sample (Barger et al.\ 1999c;
Ivison et al.\ 1998) discussed in section 3 indicate that at least
20\ per cent of the sample show some AGN activity, we shall assume in the
following discussion that a substantial fraction of the submm light
arises from star formation.

Several groups (Smail et al.\ 1998; Eales et al.\ 1999; Lilly et al.\ 1999;
Trentham, Blain, \& Goldader 1999; Barger, Cowie, \& Sanders 1999) have
suggested that the submm sources are associated with major merger events
giving rise to the formation of spheroidal galaxies.  The approximate equality
of the optical and submm backgrounds supports this hypothesis; present-day
spheroidal and disk populations have roughly comparable amounts of metal
density, and thus their formation is expected to produce comparable amounts of
light (Cowie 1988).  Since the volume density of local ULIGS is very low
(approximately $10^{-6}\ {\rm h_{65}^{3}\ Mpc^{-3}}$ for objects with
bolometric luminosities above $5\times 10^{11}\ {\rm h_{65}^{-2}\ L_\odot}$
e.g. Sanders and Mirabel 1996)
it appears that the star formation rate in this population must have been much
higher in the past and have declined very steeply after $z=1$, which may also
be consistent with this interpretation.  For a cumulative source density of
$4.0\times 10^{4}\ {\rm deg}^{-2}$ required to reproduce the EBL with 1\ mJy
sources ($\langle N\rangle={\rm EBL}/\langle S\rangle$ with $\langle
S\rangle\sim 1$\ mJy) and redshifts in the $1-3$\ range, the average space
density is $5\times 10^{-3}\ {\rm h_{65}^{3}\ Mpc^{-3}}$ for a $q_0=0.5$
cosmology ($10^{-3}$ for $q_0=0.02$).  This space density is rather
insensitive to the upper cut-off on the redshift distribution, dropping by
only a factor of $\sim 2$ or 3 if we extend the volume calculation to $z=5$.
For comparison, the space density of present-day ellipticals is about
$10^{-3}\ {\rm h_{65}^{3}\ Mpc^{-3}}$ (Marzke et al.\ 1994).  Within the
still substantial uncertainty posed by the dust temperatures, the estimated
star formation rate from submm sources in the $z=1-3$ range is $\sim0.3\ {\rm
h_{65}\ M_\odot\ yr^{-1}\ Mpc^{-3}}$ for $q_0=0.5$, which is nearly an order
of magnitude higher than that observed in the optical, $\sim0.04\ {\rm h_{65}\
M_\odot\ yr^{-1}\ Mpc^{-3}}$ (e.g., Steidel et al.\ 1999) suggesting that at
these redshifts it is the submillimeter light which marks the bulk of the star
formation.

\acknowledgments

We would like to thank our collaborators Eric Richards, Dave
Sanders, Ian Smail, Rob Ivison, Andrew Blain and Jean-Paul
Kneib.

%
%


\begin{references}
\reference Barger, A.J., Cowie, L.L., Sanders, D.B., Fulton, E.,
Taniguchi, Y., Sato, Y., Kawara, K., Okuda, H.\ 1998, Nature, 394, 248

\reference Barger, A.J., Cowie, L.L., Trentham, N., Fulton, E.,
Hu, E.M., Songaila, A., Hall, D.\ 1999a, \aj, 117, 102

\reference Barger, A.J., Cowie, L.L., Sanders, D.B.\ 1999b,
\apj, 518, L5

\reference Barger, A.J., Cowie, L.L., Smail, I., Ivison, R.J.,
Blain, A.W., Kneib, J.-P.\ 1999c, \aj, in press

\reference Barger, A.J., Cowie, L.L., Richards E.A.\
1999, \aj, to be submitted

\reference Blain, A.W., Kneib, J.-P., Ivison, R.J., Smail, I.\
1999, \apj, 512, L87

\reference Carilli, C.L., Yun, M.S.\ 1999, \apj, 513, L13

\reference Condon, J.J.\ 1992, \araa, 30, 575

\reference Cowie, L.L.\ 1988, in The Post-Recombination Universe,
N.\ Kaiser \& A.N. Lasenby, Dordrecht: Kluwer, 1

\reference Cowie, L.L., Barger A.J., Richards E.A.\
1999 \aj, to be submitted

\reference Fixsen, D.J., Dwek, E., Mather, J.C., Bennett, C.L.,
Shafer, R.A.\ 1998, \apj, 508, 123

\reference Eales, S.\ et al.\ 1999, \apj, 515, 518

\reference Frayer, D.T., Ivison, R.J., Scoville, N.Z., Yun, M.,
Evans, A.S., Smail, I., Blain, A.W., Kneib, J.-P.\ 1998, \apj, 506, L7

\reference Frayer, D.T., Ivison, R.J., Scoville, N.Z., Evans, A.S.,
Yun, M., Smail, I., Barger, A.J., Blain, A.W., Kneib, J.-P.\ 1999,
\apj, 514, 13L

\reference
Genzel, R. et al.\ 1998, \apj, 498, 579

\reference Guiderdoni, B., Bouchet, F.R., Puget, J.-L.,
Lagache, G., Hivon, E.\ 1997, Nature, 390, 257

\reference Hauser, M.G.\ et al.\ 1998, 508, 25

\reference Holland, W.S.\ et al.\ 1999, \mnras, 303, 659

\reference Hughes, D.H. et al.\ 1998, Nature, 394, 241

\reference Ivison, R., Smail, I., Le Borgne, J.-F., Blain, A.W.,
Kneib, J.-P., B\'ezecourt, J., Kerr, T.H., Davies, J.K.\ 1998,
\mnras, 298, 583

\reference Ivison, R. et al.\
1999,
\mnras, submitted.

\reference Lilly, S.J.\ et al.\ 1999, \apj, 518, 641

\reference
Marzke, R.O., Geller, M.J., Huchra, J.P., Corwin, Jr., H.G.\ 1994,
\aj, 108, 437

\reference
Pettini, M., Kellogg, M., Steidel, C.C., Dickinson, M., Adelberger, K.L.,
Giavalisco, M.\ 1998, \apj, 508, 539


\reference Puget, J.-L., Abergel, A., Bernard, J.-P., Boulanger, F.,
Burton, W.B., Desert, F.-X., Hartmann, D.\ 1996, A\&A, 308, L5

\reference Richards, E.A., Kellermann, K.I., Fomalont, E.B.,
Windhorst, R.A., Partridge, R.B.\ 1998, \aj, 116, 1039

\reference Richards, E.A.\ 1999a, \apj, 513, 9L

\reference Richards, E.A.\ 1999b, \apj, in press

\reference Richards, E.A., Fomalont, E.B., Kellermann, K.I.,
Partridge, R.B., Windhorst, R.A., Cowie, L.L, Barger, A.J.\ 1999c,
\apj, submitted

\reference Rigopoulou, D., Lawrence, A., Rowan-Robinson, M.\ 1996,
\mnras, 278, 1049

\reference Sanders, D.B., Mirabel, I.F.\ 1996, \araa, 34, 749

\reference Schlegel, D.J., Finkbeiner, D.P., Davis, M.\ 1998,
\apj, 500, 525

\reference Scoville, N., Young, J.S. 1983, 
\apj, 265, 148

\reference Smail, I., Ivison, R.J., Blain, A.W.\ 1997, \apj, 490, L5

\reference Smail, I., Ivison, R.J., Blain, A.W., Kneib, J.-P.\ 1998,
\apj, 507, 21L

\reference Smail, I., Ivison, R.J., Kneib, J.-P., Cowie, L.L.,
Blain, A.W., Barger, A.J., Owen, F.N., Morrison, G.E.\ 1999, \mnras, in press,
[astro-ph/9905246]

\reference
Steidel, C.C., Adelberger, K.L., Giavalisco, M., Dickinson, M., Pettini, M.\
1999, [astro-ph/9811399]

\reference Thronson, H.A., Telesco, C.M. 1986, \apj, 311, 98

\reference
Trentham, N., Blain, A.W., Goldader, J.\ 1999, \mnras, 305, 61 

\reference Windhorst, R.A., Fomalont, E.B., Kellermann, K.I.,
Partridge, R.B., Richards, E., Franklin, B.E., Pascarelle, S.M.,
Griffiths, R.E.\ 1995, Nature, 375, 471

\end{references}
\end{document}